\documentclass[3p,twocolumn]{elsarticle}
\begin{document}

\begin{frontmatter}

\title{Systematic trends in beta-delayed particle emitting nuclei: 
The case of $\beta$p$\alpha$ emission from $^{21}$Mg}

\author[aar]{M.V. Lund\corref{cor1}}
\ead{mvl07@phys.au.dk}
\author[cern,iem]{M.J.G. Borge}
\author[iem]{J.A. Briz} 
\author[lund]{J. Cederk\"{a}ll}
\author[aar]{H.O.U. Fynbo} 
\author[aar]{J. H. Jensen} 
\author[got]{B. Jonson} 
\author[aar]{K. L. Laursen}
\author[got]{T. Nilsson} 
\author[iem]{A. Perea}
\author[iem]{V. Pesudo}
\author[aar]{K. Riisager}
\author[iem]{O. Tengblad} 
\cortext[cor1]{Corresponding author}

\address[aar]{Department of Physics and Astronomy, Aarhus University,
 DK--8000, Aarhus C, Denmark}

\address[cern]{ISOLDE, PH Department, CERN, CH--1211 Geneve 23, Switzerland  }

\address[iem]{Instituto de Estructura de la Materia, CSIC, E-28006 Madrid, Spain}

\address[lund]{Department of Nuclear Physics, Lund University, SE-221 00 Lund, Sweden}

\address[got]{Fundamental Fysik, Chalmers Tekniska H\"{o}gskola, SE--41296
  G\"{o}teborg, Sweden}

\begin{abstract}
  We have observed $\beta^+$-delayed $\alpha$ and p$\alpha$ emission
  from the proton-rich nucleus $^{21}$Mg produced at the ISOLDE
  facility at CERN. The assignments were cross-checked with a time
  distribution analysis. This is the third identified case of
  $\beta$p$\alpha$ emission. We discuss the systematic of beta-delayed
  particle emission decays, show that our observed decays fit
  naturally into the existing pattern, and argue that the patterns are
  to a large extent caused by odd-even effects.
\end{abstract}


\end{frontmatter}

\section{Introduction}
Beta-delayed particle emission is an important decay mode for exotic
nuclei and allows many aspects of nuclear structure to be probed, see
the two recent reviews \cite{bib:Borge08,bib:Pfutzner} for a
comprehensive overview. We report here the first observation of
$\beta\alpha$ emission as well as the rare $\beta$p$\alpha$ emission
from the nucleus $^{21}$Mg. Based on these observations we have
identified systematic patterns in the occurence of beta-delayed
particle decays in proton-rich nuclei. We shall present and discuss
these as well.

A detailed description of beta-delayed particle emission must include
consideration of local nuclear structure effects, but its occurence is
in general dominated by the available energy, i.e.\ the difference
between the $Q_{\beta}$-value and the particle separation energy. As
is well known, for an isobaric chain with mass number $A$ the
$Q_{\beta}$ values will increase and the proton and neutron separation
energies decrease as one moves from the beta stability line towards
the driplines (modulated for even $A$ by the pairing term). The
$\alpha$ particle separation energy tends for light nuclei to be
minimal for $N=Z$ nuclei, but the minimum moves towards more proton
rich nuclei and reaches the proton dripline at $A \sim 50$. 
This causes a clear pattern for beta-delayed multi-particle emission,
with $\beta$2p and $\beta$3p taking place close to the proton
dripline, $\beta$2n, $\beta$3n etc starting from about halfway to the neutron
dripline, while $\beta$2$\alpha$ is seen from $A=8,9,12$ nuclei close
to stability. (To the extent that these decays are sequential one can
of course regard them as $\beta\alpha$ decays to the unstable $A=5,8$
nuclei.) Similar patterns appear in beta-delayed single-particle
emission although exceptions occur for the very light nuclei such as
the large $P_n$ values for $N=10$ nuclei and the $\beta\alpha$
emission from neutron-rich N-isotopes.

We focus first on the multi-particle $\beta$p$\alpha$ decay and
return in the discussion to the general patterns of beta-delayed
particle emission.

\section{Experimental results}
The $^{21}$Mg activity was produced at the ISOLDE facility at CERN by
a 1.4 GeV proton beam impinging upon a SiC target. The produced atoms
were extracted, laser ionized, accelerated to 60 keV, led through a
mass separator into the experimental set-up, and implanted in the
window of a gas-Si telescope opposed by a Si(DSSSD)-Si telescope.
A full account of the experimental procedure is given in
\cite{bib:MLu15}. The collected source also contained a substantial
amount of $^{21}$Na.

\begin{figure}
%  \vspace{2cm}
\resizebox{0.50\textwidth}{!}{
  \includegraphics{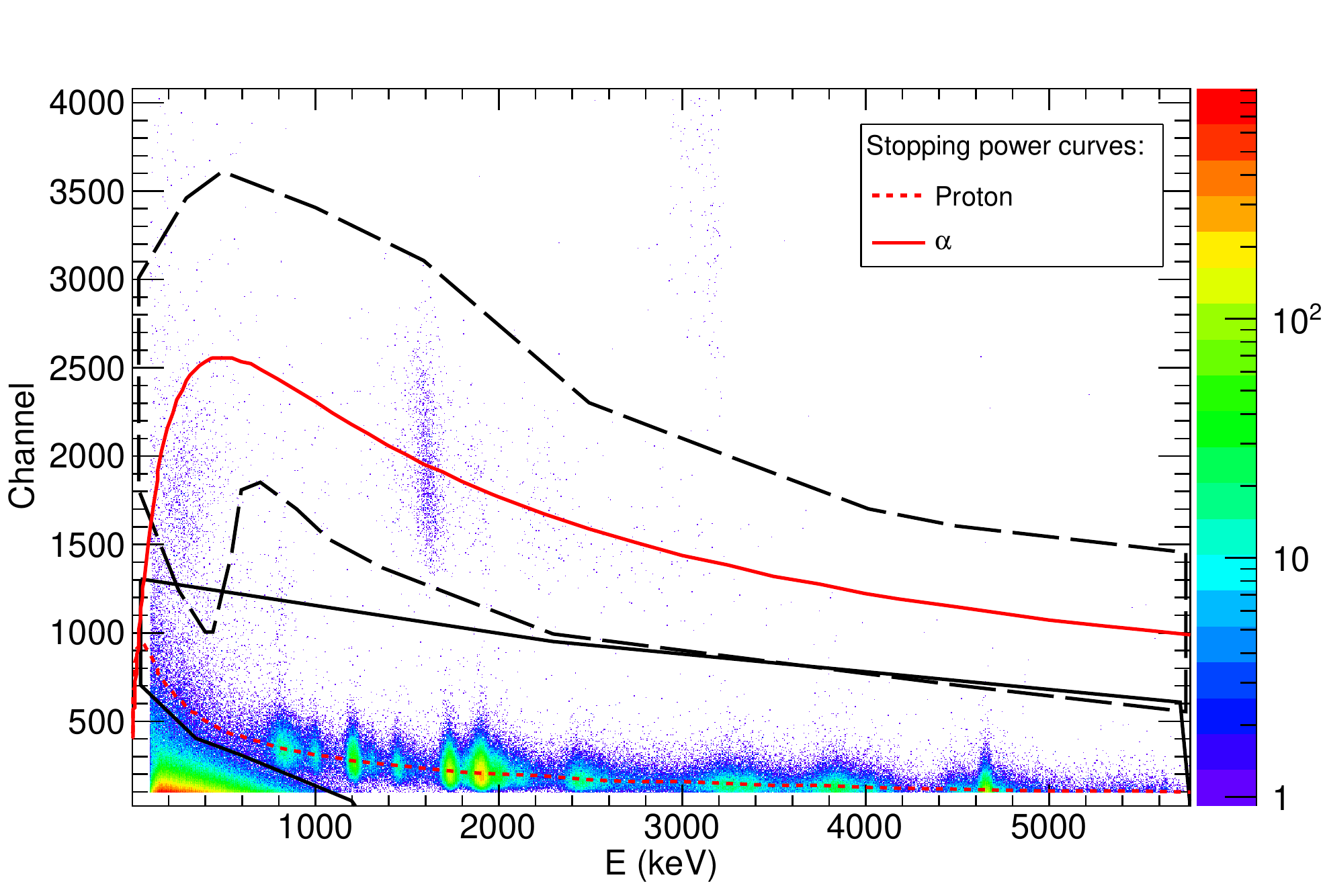}
}
\caption{(Color online) $\Delta E$-$E$ plot from the Gas-Si
  telescope with the gas channel on the vertical axis and the
  deposited energy in the silicon detector on the horizontal
  axis. The scaled stopping powers for $\alpha$'s and protons
  are shown on top of the data in solid and dashed red,
  respectively. The graphical cut used for the $\alpha$-particles is
  shown with the dashed black closed line and the cut for the protons
  is shown by the solid black closed line. The events with 3.18 MeV
  in the silicon detector and high energy deposition in the gas are
  due to a contamination of $^{148}$Gd.}
\label{fig:BananaPlot}
\end{figure}

The data from the Gas-Si charged particle telescope are presented as a
$\Delta E$-$E$ spectrum in Fig.\ \ref{fig:BananaPlot}.  Rescaled
stopping powers \cite{bib:SRIM} for $\alpha$ particles and protons
(evaluated for silicon, but representing the total energy loss in the
collection foil, the gas detector and the Si dead layer) are drawn in
the figure and match the data well, indicating the presence of
$\beta\alpha$ decays on top of the previously established
\cite{bib:Sextro,bib:JCThomas} $\beta p$. The $\beta$-particle
component in the lower left corner of Fig.\ \ref{fig:BananaPlot}
overlaps with protons below 1150 keV and $\alpha$-particles below
700 keV making particle identification difficult at low energy.
The $\alpha$-particles are stopped in the DSSSD and cannot be
separated there from the more intense proton branches.

\begin{figure}
\resizebox{0.50\textwidth}{!}{
  \includegraphics{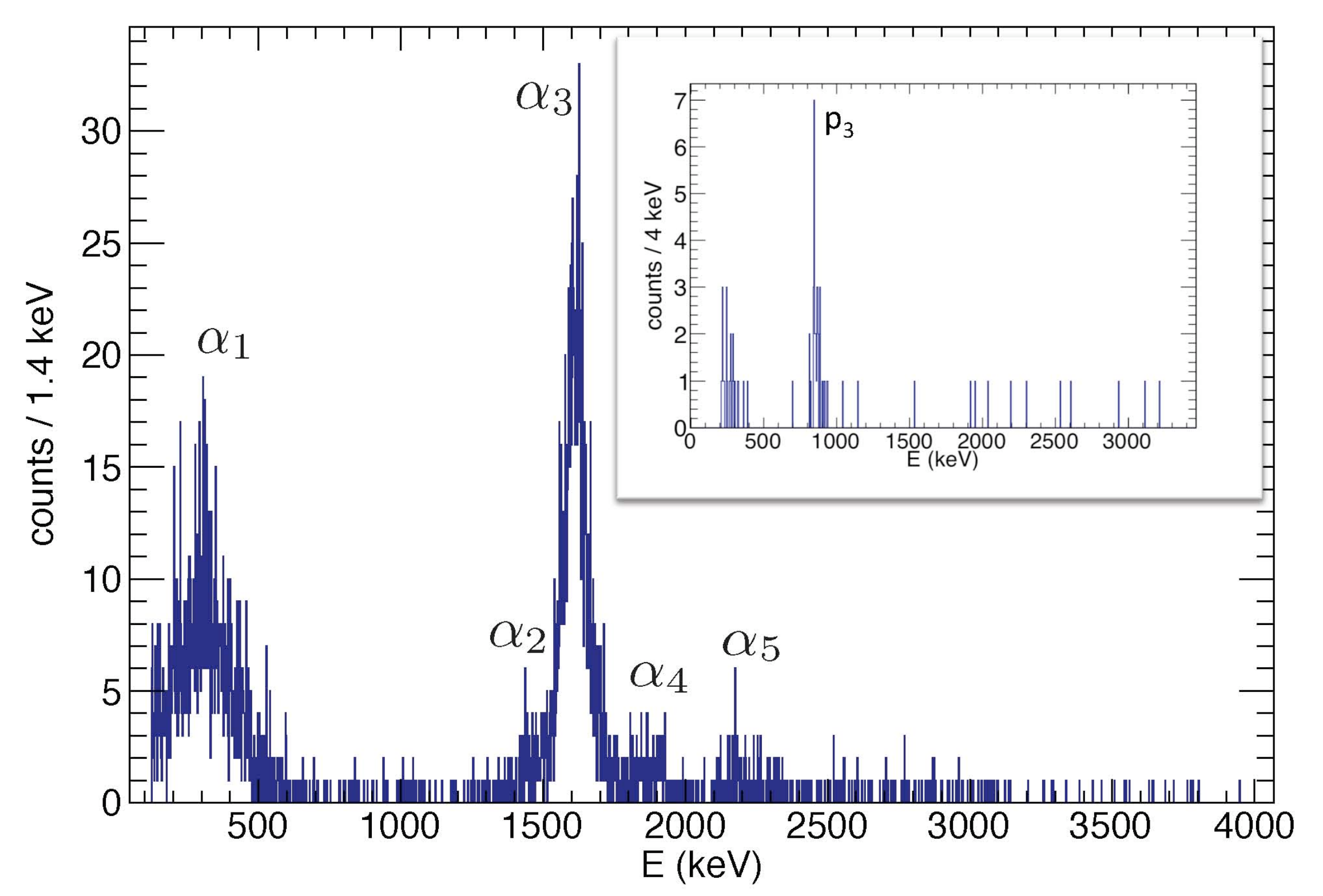}
}
\caption{(Color online) Singles $\alpha$ spectrum extracted from the Gas-Si
  telescope. The inset in the top
  right corner shows the DSSSD proton spectrum in
  coincidence with the low energy $\alpha_1$ line.}
\label{fig:AlphaSpectrum}
\end{figure}

The $\alpha$-particle spectrum extracted from the gas-telescope by
applying the gate drawn as a dashed black closed line in Fig.\
\ref{fig:BananaPlot} is shown in Fig.\ \ref{fig:AlphaSpectrum}.  Apart
from a remaining background component at low energy five $\alpha$
branches can be identified in the spectrum.  The $\alpha_2$,
$\alpha_3$, $\alpha_4$ and $\alpha_5$ lines naturally fit into the
$^{21}$Mg scheme put forward in \cite{bib:MLu15} as $\beta$-delayed
single $\alpha$ branches (see \cite{bib:MLu15} for the full decay
scheme).  The $\alpha_1$ line, with measured laboratory energy 714(12)
keV, does not fit with a transition between known levels in $^{21}$Na
and $^{17}$F. However, it does agree with a known $\alpha$-particle
transition from $^{20}$Ne to $^{16}$O observed in the decay of
$^{20}$Na \cite{bib:Kasper_20Na} with a laboratory energy of 714(4)
keV.

A conclusive particle identification for $\alpha_1$ was not possible
from the $\Delta E$-$E$ plot, but strong support for the above
assignment comes from the observation of a coincident line in the
DSSSD detector, assigned to be the preceding proton. This proton
branch p$_3$ (the numbering is chosen to be consistent with the full
data set discussed in \cite{bib:MLu15}) is displayed as the inset in
Fig.\ \ref{fig:AlphaSpectrum}. From the measured energy we deduce
$E_{\mathrm{cm}}$(p$_3$) = 919(18) keV which leads to the
interpretation of $\alpha_1$ and p$_3$ as being due to
$\beta$p$\alpha$ decay of $^{21}$Mg through the $5/2^+$ isobaric
analogue state (IAS) at 8.975 MeV in $^{21}$Na via proton emission to
the 5.621 MeV $3^-$ level in $^{20}$Ne and finally $\alpha$ emission
to the ground state of $^{16}$O. The total branching ratio of this
decay branch is found to be $1.6(3) \cdot 10^{-4}$.  This proton
branch from the IAS has not been observed earlier and
$\alpha$-emission from excited states of $^{21}$Na have only been
reported in one earlier experiment \cite{bib:Gru77}.

\subsection{Time distribution analysis}
As mentioned above our data are contaminated by $^{21}$Na, other
small contaminants could in principle also be present. The observed
$\beta\alpha$ and $\beta$p$\alpha$ branches are quite weak, so a
cross-check of the assignment is valuable. This is done by
considering the time distribution of the events.

Several factors influence the time distribution of the recorded
$^{21}$Mg events, see \cite{bib:MLu15} for an exhaustive
discussion. We shall use as reference the experimental
time distribution recorded for events within the proton gate in Figure
\ref{fig:BananaPlot} and with energy above 1150 keV. The energy gate
ensures that the reference distribution only contains protons from the
decay of $^{21}$Mg. 
%and the fact that the proton beam
%on the ISOLDE target is pulsed with pulse distance a multiple of 1200
%ms, provide a natural way of distinguishing between the two
%decays. The timescale for Mg ions to diffuse out of the target, be
%ionized and transported to the setup is of the order of 100 ms so the beam
%was let into the setup during the first 300 ms following proton impact
%on target. The resulting time distribution of the two decays is
%therefore somewhat complex. 
The halflives of $^{21}$Mg and $^{21}$Na, 122(2) ms and 22.49(4) s
\cite{bib:Nubase}, differ greatly as do the corresponding time
distributions. Other contaminants are also expected to differ from $^{21}$Mg.
%We note
%that this test will also be able to discriminate against contaminants
%from other activities appearing in the ion beam.
 
Some of the $\beta\alpha$ branches have quite low statistics and we
therefore compare their time distribution directly to the reference
distribution.  This can be done very efficiently with the empirical
distribution function (EDF) statistics \cite{bib:Ste86} that give
powerful goodness-of-fit tests. The basic principle is to compare the
shape of the data sample to the reference shape by measuring the
distance between the two cumulated distributions. For experimental and
reference distributions with values $EDF_i$ and $F_i$ in bin $i$, the
three most frequently used EDF statistics are \cite{bib:Cho94}
Kolmogorov-Smirnov
\[D = \sqrt{N} \max_i |EDF_i-F_i|,\] 
Cramer-Von Mises 
\[W^2 = N \sum_i (EDF_i-F_i)^2 p_i \]
and Anderson-Darling 
\[A^2 = N \sum_i \frac{(EDF_i-F_i)^2 p_i}{F_i(1-F_i)}, \] 
where $N$ is the total number of counts and $p_i$ is the probability
to be in bin $i$ in the reference distribution.  The second column of
Table \ref{tab:TimeResult} gives the 95\% confidence levels for the
three EDF statistics obtained through Monte Carlo simulation, values
below these levels indicate the time distribution for the different
lines are consistent with the one of $^{21}$Mg.  More details on the
confidence levels are given in \cite{bib:MLu15}.

\begin{table}
\caption{EDF goodness-of-fit tests of the time distribution of the
  individual observed lines. The first column denotes the test, the
  second column gives the 95\% confidence level (obtained through
  Monte Carlo simulations) for having the $^{21}$Mg time distribution.}
\label{tab:TimeResult} 
\centering
\begin{tabular}{cc|ccccc}
\hline\noalign{\smallskip}
 & 95\% c.l. & p$_3$ & $\alpha_1$ & $\alpha_{2}, \alpha_3$ & $\alpha_4$ & $\alpha_5$ \\
\noalign{\smallskip}\hline\noalign{\smallskip}
D & 1.31 & 1.22 & 1.64 & 0.64 & 1.08 & 0.81 \\
W$^2$ & 0.46 & 0.33 & 0.41 & 0.05 & 0.33 & 0.12 \\
A$^2$ & 2.49 & 1.46 & 3.61 & 0.64 & 1.73 & 0.80 \\
\noalign{\smallskip}\hline
\end{tabular}
\end{table}

The EDF test results in Table \ref{tab:TimeResult} show that all
lines, except for $\alpha_1$, agree with the reference distribution.
The agreement is particularly good for the strongest line, $\alpha_3$.
As mentioned above there is a contamination of $\beta$-particles in
$\alpha_1$ that come from both $^{21}$Na and $^{21}$Mg. We would
therefore expect the time distribution for $\alpha_1$ to be mainly
that of $^{21}$Mg with a small component of $^{21}$Na. The EDF tests
are sufficiently sensitive to see the effect of the small $^{21}$Na
contribution. We expect, and do observe, that the upper part of the $\alpha_1$
distribution has smaller contamination level. The fact that the
coincident p$_3$ distribution is consistent with being from $^{21}$Mg
implies that we can safely assign the $\beta$p$\alpha$ transition, as
well as all $\beta\alpha$ transitions, to the decay of $^{21}$Mg.

\section{Discussion}

\subsection{Other $\beta$p$\alpha$ cases}
The $\beta$p$\alpha$ decay mode is very rare as described in
\cite{bib:Borge08} with only two previously established cases: $^9$C
and $^{17}$Ne. For two further candidates, $^{13}$O and $^{23}$Si, the
decay mode has not been seen so far. Most searches have concentrated
on seeing particle emission from the IAS in the beta-daughter due to
the large beta-strength to this state.

The case of $^9$C is special in that all states populated in the
beta-daughter $^9$B break-up into two $\alpha$-particles and a proton,
see \cite{bib:C9} and references therein. This could be presented as a
100\% branching ratio for $\beta$p$\alpha$ or $\beta\alpha$p decay to
$^4$He, but the decays of the $A=9$ nuclei are special in several
aspects \cite{bib:Borge08,bib:Pfutzner} and are not typical for this
decay mode.

Although $\beta$p$\alpha$ has not been observed so far for $^{13}$O it
must occur since $\beta$-decays to the IAS in $^{13}$N have been observed
\cite{bib:O13} and close to half of the IAS decays are known from
reaction experiments \cite{bib:TUNL} to go via proton-emission to
$\alpha$ unbound states in $^{12}$C or $\alpha$-emission to proton
unbound states in $^9$B. Actually, the final state in both cases will
be a proton and three $\alpha$-particles which makes the decay more
challenging to observe. The total branching ratio for the decay mode
can be estimated to be $0.9(3) \cdot 10^{-4}$.

For $^{17}$Ne both decay orderings, $\beta$p$\alpha$ and
$\beta\alpha$p, have been observed \cite{bib:Ne17} with a total
branching ratio for the decay mode of $1.6(4) \cdot 10^{-4}$. All
observed decays proceed through the IAS in $^{17}$F and go to the
final nucleus $^{12}$C.

Adding now our observation of $^{21}$Mg($\beta$p$\alpha$)$^{16}$O it
is striking that all cases go through an $\alpha$-conjugate
%($x\alpha$) 
nucleus, namely $^8$Be, $^{12}$C, $^{16}$O and $^{20}$Ne
respectively. Before drawing any firm structure conclusions we
shall consider the broader systematics of beta-delayed particle
emission in $Z>N$ nuclei.

\subsection{Systematics of beta-delayed decays}
\begin{figure}
\resizebox{0.40\textwidth}{!}{
  \includegraphics{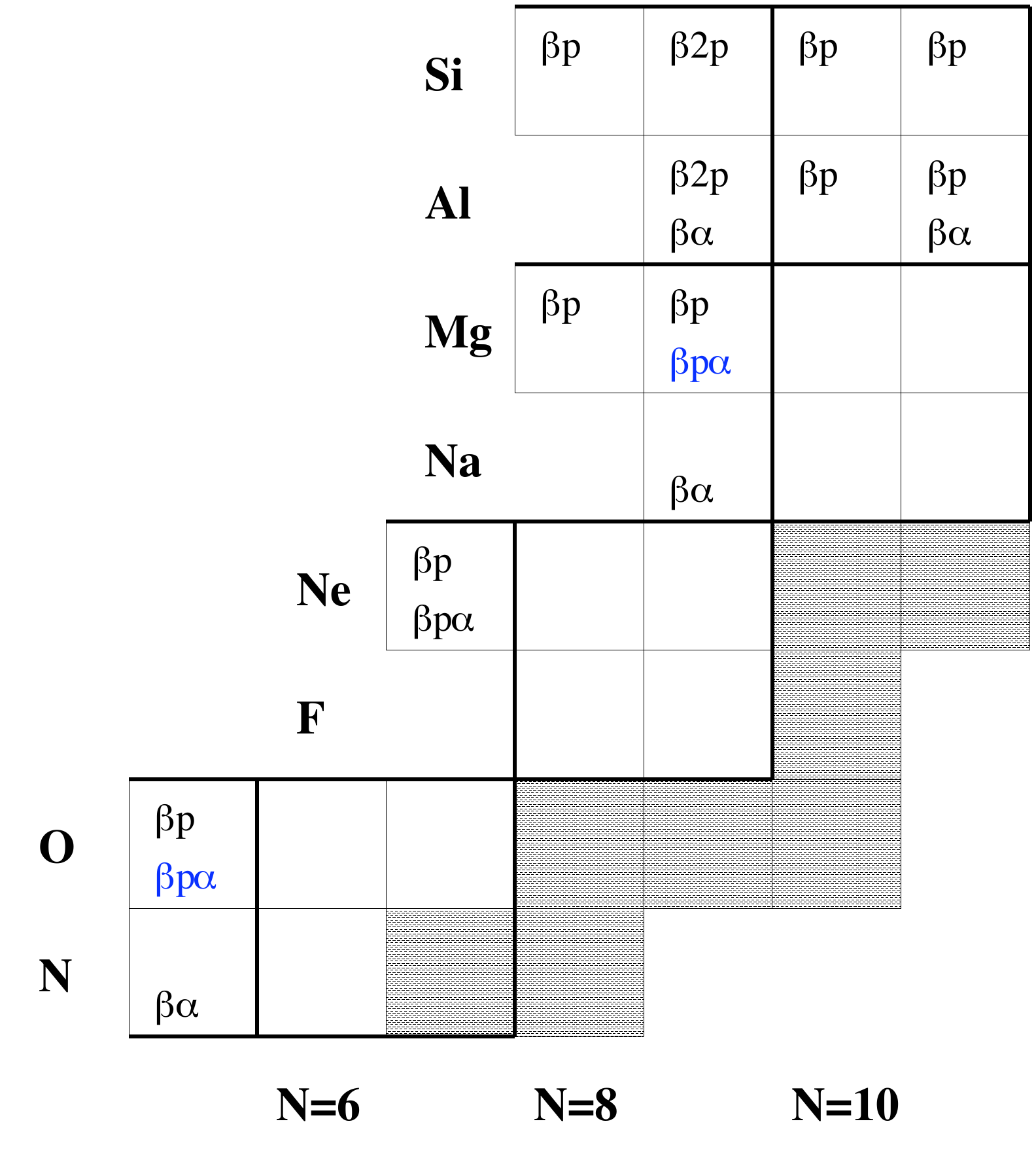}
}
\caption{(Color online) The $\beta^+$-decaying isotopes of the
  elements from N to Si. Dark squares indicate stable isotopes. The
  experimentally observed beta-delayed particle decay modes are
  indicated, the $\beta$p$\alpha$ decay modes for $^{21}$Mg (seen for
  the first time here) and $^{13}$O (see section 3.1) are marked in blue.}
\label{fig:ton}
\end{figure}

Similar patterns also appear in other beta-delayed particle decays
(see \cite{bib:Borge08,bib:Nubase} for more data and for
references to the original work). One closely related example is
$\beta\alpha$ decays that occur for all bound $A=4n$, $T_z = -1$
nuclei up to $A=40$: $^8$B, $^{12}$N, $^{20}$Na, $^{24}$Al, $^{28}$P,
$^{32}$Cl, $^{36}$K and $^{40}$Sc.
%, and has apart from the
%$\beta$p$\alpha$ cases $^{17}$Ne and $^{21}$Mg only been reported
%otherwise from $^{22}$Al in this mass region. 
The $\beta$p decays are
well established \cite{bib:Borge08} to occur strongly in $A=4n+1$,
$T_z=-3/2$ nuclei. The $\beta$2p decays of $^{22}$Al and $^{26}$P
and the $\beta$3p decay of $^{31}$Ar \cite{bib:Ar31} also all end up in
an $\alpha$-conjugate nucleus. The decays observed for the elements N
to Si are shown in figure \ref{fig:ton}. Note that the $\beta\alpha$
and $\beta$p modes are not marked explicitly when the $\beta$p$\alpha$
or $\beta$2p modes also occur.

\begin{figure}
\resizebox{0.50\textwidth}{!}{
  \includegraphics{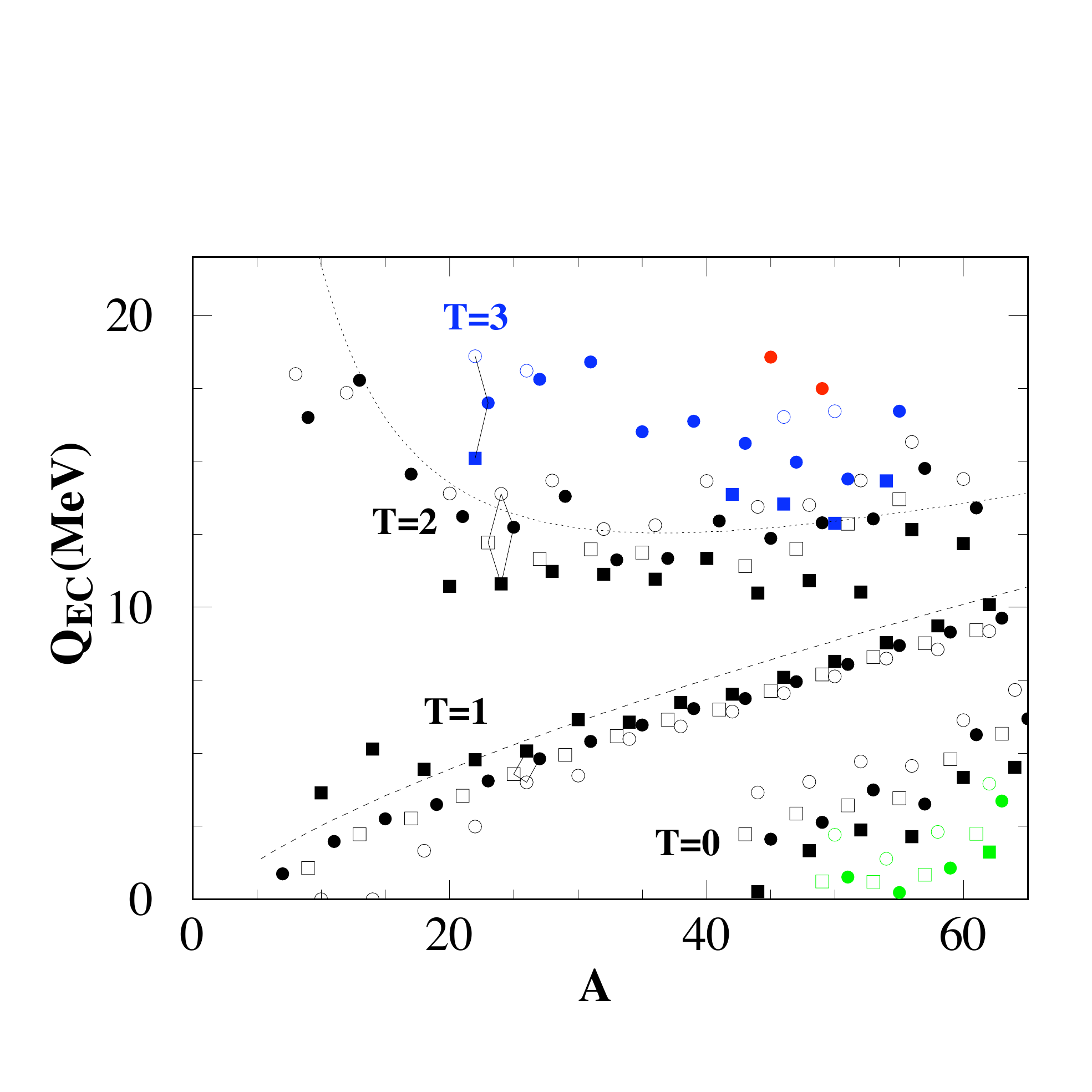}
}
\caption{(Color online) $Q_{EC}$ values for nuclei up to Ge. Filled
  symbols indicate nuclei with even $Z$, open symbols odd
  $Z$. Squares indicate nuclei with even $N$, circles odd $N$. The
  isospin values given are for the even-even nuclei. The green and red
symbols correspond to $T=1$ ($T_z=1$) and $T=4$. The dashed (dotted) line is a
liquid drop estimate of $Q_{EC}$ for nuclei with $T_z = -1/2$
($-3/2$). The full lines indicate quartets of Al-Si nuclei.}
\label{fig:Qsys}
\end{figure}

In the following we shall argue that the observed patterns are likely
(except for the very lightest nuclei) to be related to odd-even effects
rather than $\alpha$-cluster structure.  We start by considering the
systematics of $Q_{EC}$-values for nuclei with $Z>N$ as
illustrated in figure \ref{fig:Qsys}. Even though many effects
contribute to the masses in this region, a liquid drop estimate
reproduces the trend of $Q_{EC}$ for the odd-A nuclei with $T_z =
-1/2$ (dashed line) where only the Coulomb term enters, as well as for
$T_z = -3/2$ (dotted line) where the asymmetry term also
contributes. Note that $Q_{EC}$ in the latter case varies little for
$A$ between 25 and 50.

\begin{figure}
\resizebox{0.47\textwidth}{!}{
  \includegraphics{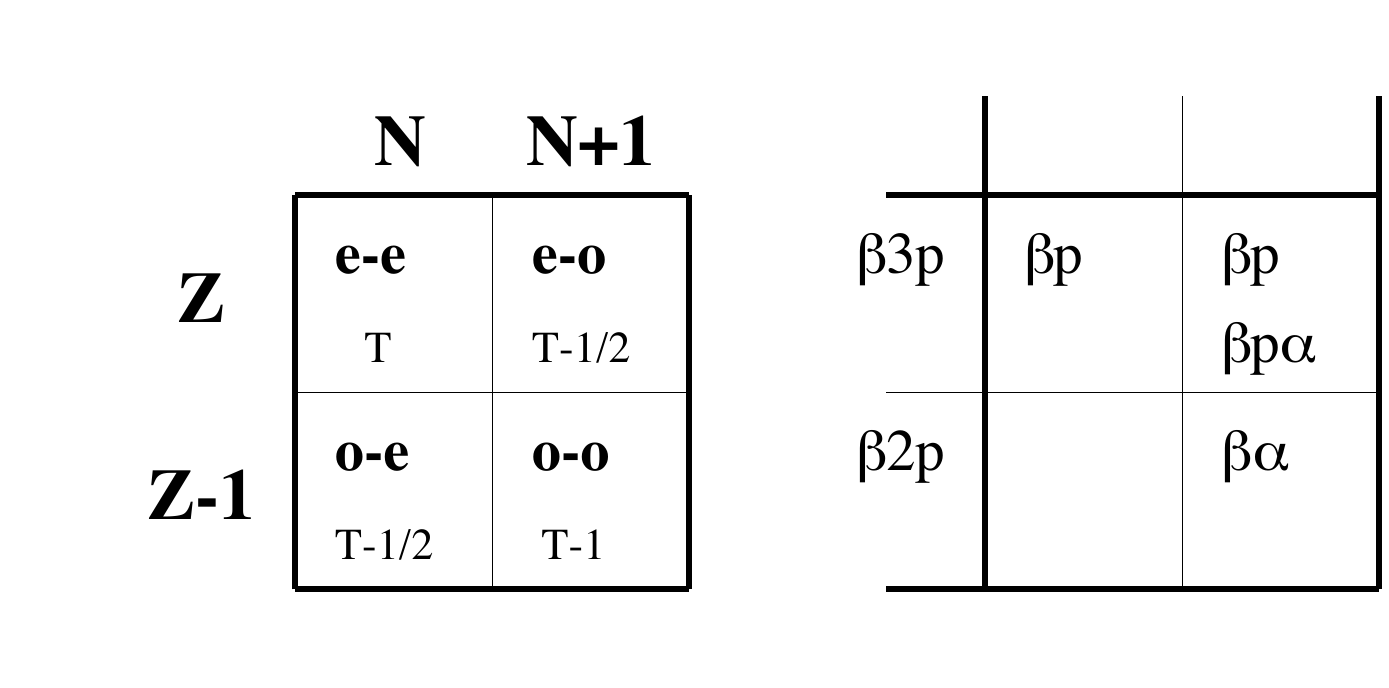}
}
\caption{Left part: The quartet of nuclei with similar $Q$-value.
  Right part: the favoured beta-delayed decay modes. See the text for
  details.}
\label{fig:sketch}
\end{figure}

The experimental data show that the $Q_{EC}$-values are roughly the same
for each ``quartet'' of four nuclei that, as illustrated in the left
panel of figure \ref{fig:sketch}, have proton and neutron numbers
(Z,N), (Z,N+1), (Z-1,N), (Z-1,N+1) where both Z and N are even. This
is pronounced for quartets where the even-even nucleus has $T=1$, and
holds to a lesser degree also for $T=2$ for mass numbers up to 40. The
reason for this is that the two odd-A nuclei are at the same distance
from the beta-stability line and therefore have about the same
$Q$-value, as also shown by the liquid drop estimate. Without a
pairing term in the liquid drop formula the $Q$-value for the even-even
nucleus would be larger and the odd-odd smaller, but the odd-even effects
counteracts this and as can be seen from figure \ref{fig:Qsys} the
magnitudes are even reversed for most nuclei. For
the quartet with $T=1$ (and $T_z = -1$) the odd-odd nucleus has $N=Z$
and is therefore extra bound, this happens to result in $Q$-values
that are almost the same for all four nuclei. The quartets are
indicated in figure \ref{fig:ton} by thicker lines.

The observed decay patterns now follow from the energetics and are
illustrated in the right panel of figure \ref{fig:sketch}. The
$\beta\alpha$ decays should occur in odd-odd nuclei, since they have
slightly higher $Q$-values and the daughter alpha particle separation
energies tend to be smallest here. The $\beta$p decays need low proton
separation energies in the daughter nucleus and therefore are more
prominent for even Z, starting (as one goes from stability towards the
proton dripline) in an even-odd nucleus. The $\beta$p$\alpha$ decay
should be favoured in even-odd nuclei, and $\beta$2p and $\beta$3p
decays should occur in odd-odd and even-odd nuclei, respectively, by
extending these arguments.

Experimentally, the $\beta\alpha$, $\beta$p and $\beta$p$\alpha$ decays
appear first in the quartets where the even-even nucleus has $T=2$ and
the odd-odd nucleus $T=1$, but $\beta$p occurs also in $^{59}$Zn,
$^{65}$Ge and heavier nuclei. The beta-delayed multi-proton decays
appear in more exotic nuclei, but it is noteworthy that $\beta\alpha$
in these nuclei only has been observed in the odd-odd $^{22}$Al.

Similar patterns can be expected for $\beta^-$-delayed particle
decays, although the grouping of $Q$-values is less pronounced here. The
$\beta\alpha$ and $\beta$n$\alpha$ decay modes will in general occur
further away from the beta-stability line. 

\section{Conclusion}

Our study of the decay of $^{21}$Mg has given the first evidence for
the occurence of the $\beta\alpha$ and $\beta$p$\alpha$ decay modes in
this nucleus. The assignment of these decay modes to $^{21}$Mg has
been verified through statistical tests of the time distribution of
the events. The occurence of these decay modes in the $\beta^+$ decay
of $^{21}$Mg fits naturally into the systematics of previously
observed $\beta^+$-delayed decays. We presented a brief overview of
this systematics and argued that it can be explained by the variation
in decay energy due to odd-even effects and that there is no need to
invoke specific structure effects such as alpha-clustering in spite of
$\alpha$-conjugate nuclei occuring often as final state nuclei.

This interpretation can be tested when new instances of these exotic
decays are discovered.
The $\beta$p$\alpha$ decay mode may not occur in heavier $T_z = -3/2$
nuclei than $^{21}$Mg (the $Q$-value becomes more than 10 MeV in
$^{61}$Ge, but the Coulomb barrier for $\alpha$-particle emission is
substantial then), but may be found also in the $T_z = -5/2$ nuclei
$^{23}$Si, $^{27}$S, $^{31}$Ar etc. If found in $^{20}$Mg it may help
to quantify the $^{15}$O($\alpha$,$\gamma$)$^{19}$Ne reaction rate
\cite{bib:WrePoS}. A general overview of which energetically allowed
decays have not yet been observed was given in \cite{bib:Borge08}.

\section*{Acknowledgements}
We acknowledge support from the European Union
Seventh Framework through ENSAR (contract no. 262010) and by the
Spanish research agency under number FPA2012-32443.
We thank Oliver Kirsebom for discussions.

\end{document}